\documentclass[a4paper,11pt]{article}
\usepackage{pos}
\usepackage{braket}
\usepackage[normalem]{ulem} % for \sout

\usepackage{subcaption}

\usepackage{bbold}

\usepackage[english]{babel}

\RequirePackage{color}
\definecolor{dkgreen}{rgb}{0.2,0.7,0.4}
\definecolor{dkblue}{rgb}{0.2,0.2,0.7}
\definecolor{dkred}{rgb}{0.8,0,0}
\definecolor{dkpurple}{rgb}{0.45,0.2,0.55}

\begin{document}
\title{Hilbert Space Fragmentation and Gauge Symmetry}
%% \ShortTitle{...}
\author*[a]{Thea Budde}
\author[a]{Marina Kristc Marinkovic}
\author[a]{Joao C. Pinto Barros}
\affiliation[a]{Institut für Theoretische Physik, ETH Zürich, Wolfgang-Pauli-Str. 27, 8093 Zürich, Switzerland}
\emailAdd{tbudde@phys.ethz.ch}
\abstract{
The Hamiltonian formulation of lattice gauge theories plays a central role in quantum simulations of gauge theories, and understanding their spectrum and other properties is expected to become crucial in the upcoming years. The relevant Hamiltonians in this framework possess local symmetry at each lattice site and may exhibit higher-form symmetries. There are then an exponentially large number of dynamically disconnected symmetry sectors, most of which are not translation-invariant. An exponential number of dynamically disconnected sectors, i.e., Hilbert space fragmentation, can also occur in systems in which no such symmetries have been identified. In this contribution, we describe an emergent gauge symmetry that is valid only in a subset of sectors of the fragmented $S=1$ dipole-conserving spin chain. These non-invertible symmetries can label exponentially many of the model's sectors. Simulating this Hamiltonian, which is not gauge-invariant, yields an exact quantum simulation of a gauge theory.
}
\FullConference{The 42nd International Symposium on Lattice Field Theory \\
TIFR Mumbai\\
3.11.25
}
%\tableofcontents
\maketitle

\section{Introduction}

Real-time dynamics quantities have remained a major challenge in lattice Quantum Chromodynamics (QCD) since Monte Carlo simulations are performed in Euclidean time \cite{gattringerQuantumChromodynamicsLattice2010}. This has been a major motivation to move towards a new paradigm in which quantum simulations of QCD are performed alongside (classical) Monte Carlo simulations.
Quantum technologies are being developed rapidly, promising efficient simulations of quantum many-body systems, including gauge theories \cite{Halimeh:2025vvp}. This has motivated renewed interest in the study of gauge theories in the Hamiltonian formulation with finite-dimensional local Hilbert spaces. 
 
While QCD remains a far-reaching goal, other interesting questions have emerged in simpler gauge theories. One such aspect is the thermalization of closed systems. The ergodicity of closed systems evolving in time underlies their thermalization \cite{Srednicki:1994mfb,Deutsch:2018ulr,DAlessio:2015qtq,Gogolin:2015gts}. In recent years, it has become increasingly evident that ergodicity does not hold in a wide range of systems \cite{Chandran:2022jtd,Moudgalya:2021xlu}.
Thermalization can occur if the system ergodically explores the accessible Hilbert space, subject to energy conservation and other symmetries. To verify this, ergodicity for each symmetry sector must be checked individually.
Phenomena under which only a small set of initial states break the expectations of thermalization, or in which only a subextensive fraction of the spectrum is affected, are classed as weak ergodicity breaking. Since only a small part of the spectrum is affected, this will not be visible in common statistical spectral analyses, such as level-spacing ratios. These have commonly been used to test ergodicity and to examine how the high-energy spectrum of interacting Hamiltonians can statistically resemble that of a random matrix \cite{Srednicki:1994mfb,DAlessio:2015qtq}. An example where weak ergodicity breaking occurs is when the spectrum of the Hamiltonian contains quantum many-body scars (QMBS) \cite{Moudgalya:2021xlu, Chandran:2022jtd}. 

The term QMBS encompasses multiple classes of phenomena. They were originally established in the PXP model \cite{Serbyn:2020wys, Bluvstein:2020ddz}, where a small set of eigenstates with approximately equidistant energies has unusually large overlap with two product states. Evolving these two states leads to prolonged oscillations, beyond the timescales at which thermalization is expected. The set of eigenstates is not thermal and does not fulfill the requirements of ETH. There are other models with similar phenomena \cite{desaules_prominent_2023,Pinto2026}, and there are models with non-thermal eigenstates that do not have equidistant energies and therefore do not lead to oscillations \cite{Ivanov:2025daz, Banerjee:2020tgz, Budde:2024rql}. In these cases, initial states with low entanglement evolve to values that differ from those expected under thermalization. A third class of QMBS consists of models with a small set of initial states whose dynamics will never mix with the rest of the Hilbert space, even after infinite time \cite{Moudgalya:2022nll, Pakrouski:2020hym, Moudgalya:2021xlu, Sun:2022oew}. 

In the case of strong ergodicity breaking, the system fails to thermalize from most initial states \cite{Moudgalya:2021xlu}. Beyond integrable models \cite{DAlessio:2015qtq,Vidmar2016,moudgalya_quantum_2021}, this has been observed in models with Hilbert space fragmentation \cite{Moudgalya:2021xlu, Sala_2020, Moudgalya_2022,Mukherjee:2021iki}. In the latter case, states initialized in some product state remain confined to a small subspace of the Hilbert space. Concretely, let $\ket{\psi_0}$ be a state on the Hilbert space. If the system is initialized in this state, it will be within its Krylov sector
\begin{equation}
    \mathcal{K}(\ket{\psi_0})
    \equiv \mathrm{span} \left\{ \ket{\psi_0}, H \ket{\psi_0}, H^2 \ket{\psi_0},\ ... \right\}
\end{equation}
at any future time.
Naturally, different product basis states can share the same Krylov sector, and they commonly do for generic states and interacting Hamiltonians. However, this need not be the case in general, and the Hilbert space can decompose into several distinct Krylov sectors. In an extreme case, the number of distinct Krylov sectors in the product basis can scale exponentially with system size. This is called ``Hilbert space fragmentation''. For such a product basis, the Hamiltonian becomes block-diagonal, and many of the Krylov sectors are small. One can distinguish the cases where most of the Hilbert space lies in a single Krylov sector from those where even the largest Krylov sector grows asymptotically more slowly than the dimension of the product Hilbert space with system size. The latter case is called ``strong fragmentation''. Fragmentation cannot be explained by what has been considered ``standard'' symmetries, since they can only lead to a polynomial scaling in the number of symmetry sectors \cite{Moudgalya:2021xlu}. In this context, ``standard'' symmetries are those that are generated by global sums over local densities.

In Ref.~\cite{Budde:2026fgt}, we have shown that a large class of generalized symmetries and gauge symmetry leads to exponential scaling of the number of symmetry sectors in the tensor-product basis, and therefore to fragmentation. If these symmetries are not resolved, this may be mistaken for ergodicity breaking. Non-invertible symmetries, specifically higher-form partial isometries, may be a mechanism underlying the fragmentation of other models, and we refer to \cite{Budde:2026fgt} for details.

While standard symmetries admit a unitary representation and are therefore invertible on the full Hilbert space, it has recently been shown that this is not necessary \cite{Ortiz:2025psr}. When a partial isometry $D = UP$ commutes with the Hamiltonian, transition probabilities are also conserved. Here, $U$ is unitary and $P$ is a projector onto a subspace of the Hamiltonian. Such symmetries have not yet been considered in the context of ergodicity breaking and could underlie some of the unexplained phenomena.
 
In these proceedings, we demonstrate that even when a local symmetry of the Hamiltonian cannot be identified, a fragmented model can still have a set of Krylov sectors with an emergent gauge symmetry that is not present in other Krylov sectors. This may be interpreted as non-invertible gauge symmetry. Sectors with non-invertible gauge symmetry realize the dynamics of gauge theories and can therefore be used for quantum simulations of gauge theories. 

The remainder of this contribution is organized as follows. In Section~\ref{sec:Spin1Dipole}, we introduce the so-called ``Dipole Conserving Spin Chain'', which exhibits fragmentation. We then describe how certain fragments have emergent conserved local quantities which constitute a $U(1)$ gauge symmetry. In Section~\ref{sec:non-inv}, we interpret this from the perspective of non-invertible symmetry. In Section~\ref{sec:qsim}, we demonstrate how fragmented models realize quantum simulations of gauge theories, before concluding in Section~\ref{sec:conclusion}.

\section{Local conserved quantities in the $S=1$ Dipole Conserving Spin Chain} \label{sec:Spin1Dipole}

\begin{figure}
    \centering
    \includegraphics[height=5cm, trim={4cm 0 1cm 0}, clip]{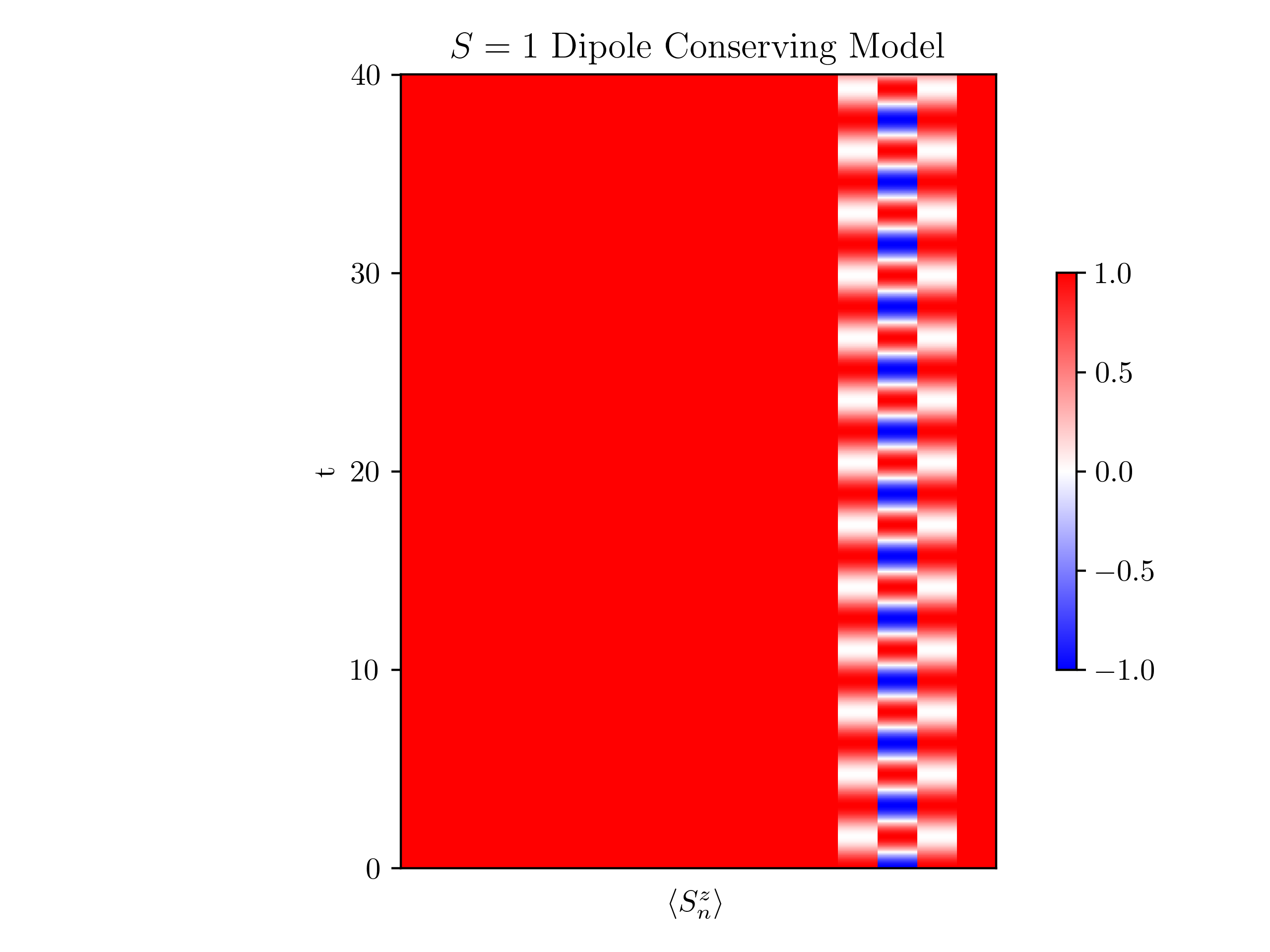}
    \includegraphics[height=5cm, trim={3cm 0 0cm 0}, clip]{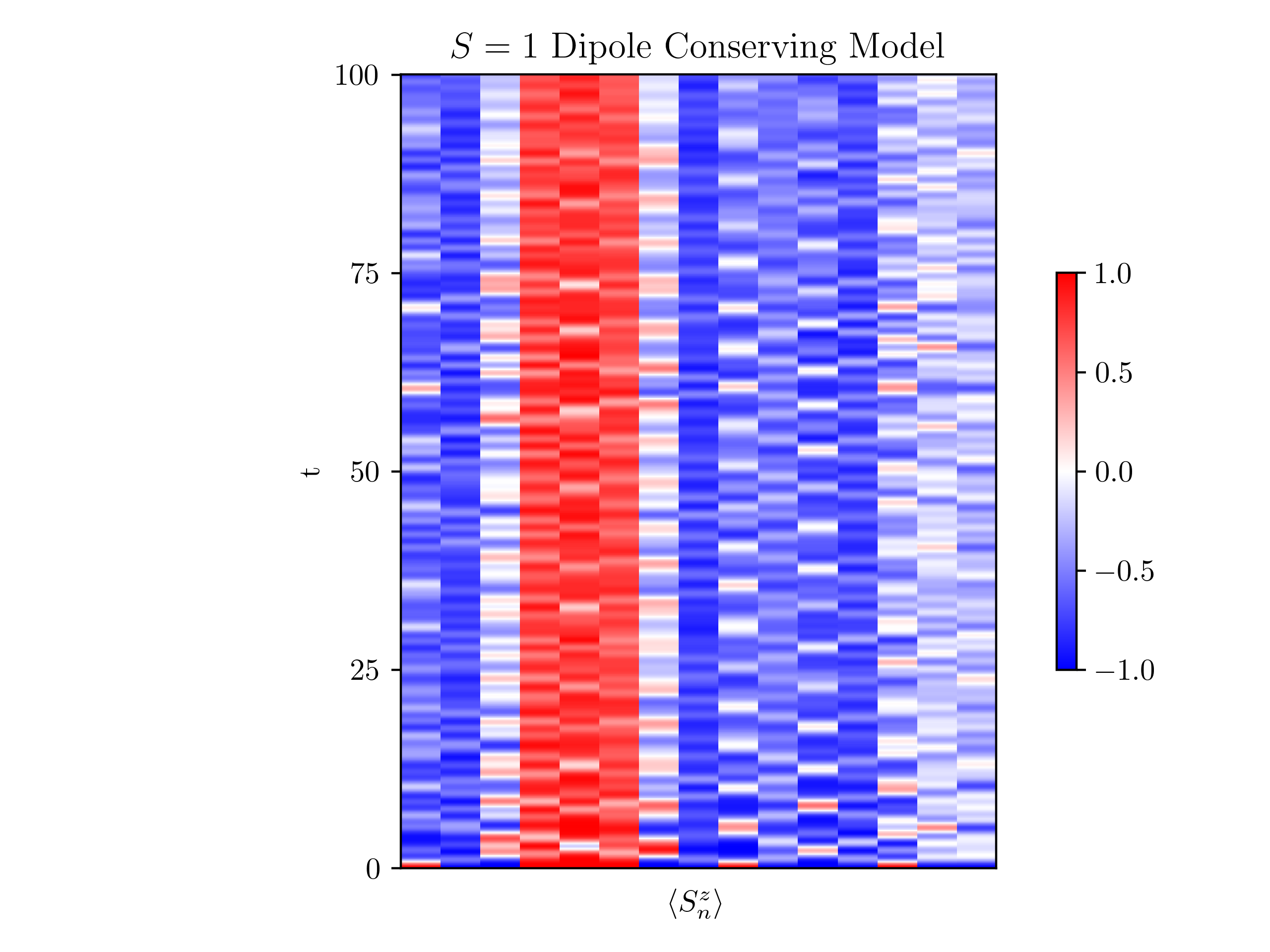}
    \caption{The real-time dynamics of two product initial states. The Krylov sectors are very small, such that significant oscillations persist even after a long time. Additionally, the system is not evolving into a translation-invariant ensemble, even though the Hamiltonian is translation invariant.}
    \label{fig:dynamics}
\end{figure}

The $S=1$ dipole conserving spin chain is a prominent example of a model with strong Hilbert space fragmentation \cite{Sala_2020, Moudgalya_2022, Moudgalya:2021xlu}.
It is given by a spin-1 chain with the Hamiltonian
\begin{equation}\label{eq:S-1Dipole}
    H = \sum_n S^+_n \left(S^-_{n+1}\right)^2 S^+_{n+2} + h.c.
\end{equation}
and we consider periodic boundary conditions, although the arguments hold for open boundaries as well. Due to the Hilbert space fragmentation, there are many states that are only dynamically connected to a very small subspace of the Hilbert space. This leads to vastly different dynamics depending on the initial state, and most states do not thermalize. Additionally, the long-time-averaged ensembles are not translation-invariant, even though the Hamiltonian is. Examples of the time evolution of some product states are depicted in Fig.~\ref{fig:dynamics}. This can lead to the same phenomenology as in disorder-free localization, as discussed in Ref.~\cite{Budde:2026fgt}.

\begin{figure}
    \begin{subfigure}[c]{0.45\columnwidth}
        \centering
        \includegraphics[width=0.6\linewidth]{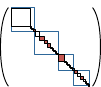}
    \end{subfigure}
    \begin{subfigure}[c]{0.55\columnwidth}
        \centering
        \resizebox{\columnwidth}{!}{
        \begin{tabular}{|c|c|c|c|c|}
            \hline
            ID & $G_{2n}$ & $\tilde{G}_{2n}$ & $G_{2n+2}$ & Compatible states $\ket{\sigma^z_{2n} \sigma^z_{2n+1} \sigma^z_{2n+2}}$ \\ \hline
            1&0 & -2 & 0 & $\ket{-0-}, \ket{+--}, \ket{--+}$  \\ \hline
            2&0 & 2 & 0 & $\ket{+0+}, \ket{++-}, \ket{-++}$ \\ \hline
            3&0 & -4 & 0 & $\ket{---}$ \\ \hline
            4&0 & 4 & 0 & $\ket{+++}$ \\ \hline
            5&1 & -3 & 0 & $\ket{0--}$ \\ \hline
            6&0 & -3 & 1 & $\ket{--0}$ \\ \hline
            7&1 & 3 & 0 & $\ket{0++}$ \\ \hline
            8&0 & 3 & 1 & $\ket{++0}$ \\ \hline
        \end{tabular}}
    \end{subfigure}
    
    \caption{Left: A sketch of the structure of the $S=1/2$ dipole conserving spin chain Hamiltonian in the product basis. The blue squares represent the global symmetry sectors. These sectors are split into many more fragments represented in black. Some sectors conserve the identified quantities eq.~\eqref{eq:conserved1}. These sectors are filled in red. Right: States for which all even sites have one of these sets of eigenvalues for eq.~\eqref{eq:conserved1}, will conserve these quantities, and they uniquely label the Krylov sector.}
    \label{fig:Hamiltonian}
\end{figure}

The model possesses two global symmetries, the magnetization and dipole moment
\begin{equation}
    M = \sum_{n=1}^L S^z_n, \quad  D = \left(\sum_{n=1}^L nS^z_n\right)\  \mathrm{mod} \ L.
\end{equation}
It has been shown that the presence of these two symmetries in an $S=1$ chain leads to exponentially many frozen states and, therefore, Hilbert space fragmentation \cite{Sala_2020}. However, the symmetries are not able to label all of the exponentially many fragments uniquely, since there can only be $\mathcal{O}(L^2)$ combinations of eigenvalues from these symmetries. The structure of the Hamiltonian is depicted in Fig.~\ref{fig:Hamiltonian}. The sectors can be labeled through non-local conserved quantities after a mapping to "clusters" and "dots" \cite{Moudgalya_2022}, but an explanation through locally generated symmetries has not been possible thus far.

We now identify local quantities that are not symmetries of the full Hamiltonian but become conserved within an exponentially large family of Krylov sectors. They are
\begin{equation}\label{eq:conserved1}
    G_n = \ket{0}\bra{0}_n, \quad \tilde{G}_n = S_n^z + 2 S^z_{n+1} + S^z_{n+2}.
\end{equation}
These local quantities do not commute with the Hamiltonian in eq.~\eqref{eq:S-1Dipole} and are therefore \emph{not} symmetries of the model. However, in certain Krylov sectors, their eigenvalues are identical throughout the sector. This means they are conserved in those Krylov sectors. 
Concretely, consider a state where all sequences of even/odd/even sites take values as listed in the Table in Fig.~\ref{fig:Hamiltonian}.
If this is the case, the even terms in the Hamiltonian will be annihilated, and the Hamiltonian effectively becomes
\begin{equation}\label{eq:Heff}
    H_{\mathrm{eff}} = \sum_{n=1}^{\frac{L}{2}} S^+_{2n-1} \left(S^-_{2n}\right)^2 S^+_{2n+1} + h.c.,
\end{equation}
which does commute with the operators in eq.~\eqref{eq:conserved1}. The same is trivially true for an analogue construction of subsequent odd/even/odd sites with the same quantum numbers. Different combinations of these eigenvalues lead to Krylov sectors of different sizes. If all triplets of eigenvalues correspond to sectors with only a single state (i.e., ID 3-8), the Krylov sector has dimension 1, and the state is frozen. If there are triplets with multiple states (i.e., ID 1 or 2), the Krylov sector will be larger and dynamical. If all triplets are of ID 1 or all of them are of ID 2, the Hilbert space maps onto the physical sector of the PXP model \cite{Sala_2020}. The sectors identified by this method are all fully connected Krylov sectors and contain no further fragments.

This insight can be used to identify additional fragments. We can find new conserved quantities by considering sequences longer than three sites. Furthermore, it is possible to use locally frozen states to disconnect parts of the lattice, forming disconnected sublattices. This leads to further fragmentation. At this point, it has not been possible to label all Krylov sectors in this way. However, the exponentially many sectors described above are uniquely labeled with this description. These sectors exist across different global symmetry sectors, as depicted in Fig.~\ref{fig:Hamiltonian}.

\section{Non-Invertible Gauge Symmetry}\label{sec:non-inv}

The quantities in eq.~\ref{eq:conserved1} are conserved only in certain sectors and are not symmetries of the Hamiltonian. But partial isometries $D_n = P G_n P$ and $\tilde{D}_n = P\tilde{G}_nP$ can be constructed, where $P$ projects on the sectors in which $G_n$ and $\tilde{G}_n$ take only the allowed values in the table of Fig.~\ref{fig:Hamiltonian} for either even or odd $n$. These operators commute with the Hamiltonian. Additionally, $P$ projects onto certain eigenvalues of local quantities, so we argue it makes sense to call $D$ and $\tilde{D}$ local. Unlike the example in Ref.~\cite{OurFragPaper}, $P$ does not project on a symmetry sector, and many fragments that cannot be explained remain. This leads us to suspect that there is a deeper, non-invertible symmetry structure in this model that has not yet been uncovered.

Similar local conserved quantities are likely to exist in other fragmented models. However, the existence of $k$-local conserved quantities can be excluded in models where all Krylov sectors are translation-invariant, e.g., in the $t-J_z$ model \cite{Moudgalya_2022}. Therefore, this fragmentation mechanism is not universal.

\section{Quantum Simulations of Gauge Theories With Fragmented Hamiltonians}\label{sec:qsim}

Advances in quantum technologies have allowed for multiple realizations of quantum simulations of gauge theories in recent years. While we are still far from being able to perform relevant quantum simulations of QCD, we are at the stage where we can dynamically exploit real-time dynamics in simpler systems. 

Quantum simulations of gauge theories are typically performed on platforms that do not themselves have gauge symmetry. The system is then tuned to a specific set of parameters, and initial states are constrained to be in a certain energy scale. Then, dynamics approximated by an effective theory emerge. This effective theory may be fragmented. This fragmentation is approximate and appears through a separation of energy scales. We note that all, or only a subset of the effective theories sectors, may have the desired gauge symmetry. As long as the system is initialized within the sectors that obey the target gauge theory, to the desired precision, it is a quantum simulation of that theory. It is not necessary for the full subspace to have gauge symmetry. 

In the following, we review various Hamiltonians prepared to simulate a toy model of choice for QCD, before proposing a new fragmented Hamiltonian that is not itself gauge-invariant. The targeted model is the $S=1/2$ Quantum Link Model (QLM) in $1+1$ dimensions \cite{Banerjee2012,Horn1981,orlandLatticeGaugeMagnets1990,QuantumLinkModels,Brower2004}. QLMs allow for maintaining an exact continuum symmetry while keeping a finite local Hilbert space. We will focus on the $U(1)$ gauge theory, which already exhibits interesting static \cite{ricoTensorNetworksLattice2014,pintobarrosMeronClusterAlgorithmsQuantum2024} and dynamical \cite{Surace_2020,desaules_prominent_2023} properties. At the same time, it also produces a prime example of a fragmented Hamiltonian. Concretely, we focus on a commonly studied model with staggered fermions, where the Hamiltonian takes the form
\begin{equation}\label{eq:QLM}
    H = J\sum_{n=1}^L \left(c_n^\dagger S^+_{n} c_{n+1} + h.c.\right) + m\sum_{n=1}^L (-1)^n c^\dagger_n c_n + \lambda \sum_{n=1}^L S^z_n.
\end{equation}
Here, $S^\pm_n$ are the $S=1/2$ raising/lowering operators and $c_n$ is a fermionic annihilation operator. The parameters $J$, $m$, and $\lambda$ parameterize the different types of gauge-invariant terms that correspond to the matter-gauge interaction, mass, and theta-term, respectively. Furthermore, we consider a lattice of size $L$ with periodic boundary conditions, identifying sites $L+1\equiv 1$. The existence of a $U(1)$ gauge symmetry is manifested by the existence of a set of generators $G_n$, which commute with the Hamiltonian. Concretely
\begin{equation}
    G_n = S^z_{n-1} + c^\dagger_n c_n - S^z_n\quad\Rightarrow\quad\left[H,G_n\right]=0.
\end{equation}
This structure guarantees that the Hilbert space can be separated into different sectors, each labeled by the set of eigenvalues of $G_n$. We can refer to these sets as $\left\{g_n\right\}_{n=1}^L$. The Hamiltonian can be diagonalized simultaneously in all sectors. The number of possible sectors grows exponentially with $L$ as the number of independent $G_n$ grows exponentially with $L$. For a more general statement, see \cite{OurFragPaper}.

The majority of the literature focuses on the so-called ``physical sector'', which is defined by
\begin{equation}
    G_n\ket{\psi}=g_n\ket{\psi},\quad g_n=\frac{(-1)^n -1}{2}.
\end{equation}
One can use the restriction to this sector to construct a concrete description that is only valid in this sector. This allows us to eliminate fermions and construct a Hamiltonian that depends only on spins \cite{Surace_2020}. When $m=\lambda=0$ and $J=1$, this gives rise to the celebrated PXP model
\begin{equation}
    H = \sum_n P_{n-1} S^x_{n} P_{n+1}.
\end{equation}
Here, $P_n$ is the projector onto the spin-down state at the $n$th site. Such a Hamiltonian is naturally implemented in experiments with Rydberg atoms, through the Rydberg blockade \cite{bernienProbingManybodyDynamics2017,Surace_2020}. The correspondence with the gauge theory sector works as long as we restrict this Hamiltonian to a Hilbert space where no up-spins can be found next to each other. Under unitary evolution, if two neighboring spins are up, they will remain up for all future times. This constitutes a rather trivial example of fragmentation.  In Ref.~\cite{OurFragPaper}, we demonstrated that the PXP model has a $\mathbb{Z}_2$ symmetry itself. Simulations of this sector are therefore also a simulation of a special $\mathbb{Z}_2$ gauge theory.

The effective Hamiltonian eq.~\eqref{eq:Heff} has been utilized to simulate this gauge theory on cold atom simulators \cite{Zhou:2021kdl}. In some contexts, it may be simpler to realize a translation-invariant version of this Hamiltonian. This would correspond to the $S=1/2$ dipole-conserving model. As long as the initial state lies in the Krylov sector corresponding to the PXP model, this also realizes a quantum simulation of a gauge theory. Simulating fragmented models that lack gauge symmetry themselves is therefore a novel option for quantum simulations of gauge theories.

\section{Conclusion}\label{sec:conclusion}

As quantum simulators of gauge theories become progressively available, we will be able to address novel questions about their dynamics. The existence of local symmetries naturally gives rise to an exponential number of disconnected sectors of the Hilbert space. While in the standard particle physics framework, we are interested in a single sector, the so-called physical sector, quantum simulators can potentially address problems where the existence of other sectors leaves an imprint on experimentally relevant questions, particularly thermalization. Assumptions of ergodicity are usually required for thermalization to occur. In fragmented systems, the Hamiltonian becomes block-diagonal in a product basis, and ergodicity is possible only when constrained to those sectors. Gauge theories provide immediate examples in which the number of blocks grows exponentially with the volume, leading to \emph{strong fragmentation}, which has received significant attention in recent years. Looking at strongly fragmented models and gauge theory through a common lens can be fruitful for both the fields of quantum simulation of gauge theories and statistical physics. In the former, it presents interesting physical phenomena that may fall outside the scope of standard numerical methods. For the latter, it provides a class of well-understood examples for which strong fragmentation occurs. 

In this work, we examined the picture above and developed a somewhat complementary perspective. Not only does gauge symmetry lead to fragmentation, but gauge symmetry can also emerge within a fragmented model without the Hamiltonian possessing gauge symmetry in the usual sense. Concretely, we demonstrated that gauge symmetry may emerge in only a subspace of the Hilbert space. We demonstrated this mechanism for the $S=1$ dipole-conserving chain. This model has been shown to break down into exponentially many disconnected sectors, yet lacks gauge symmetry. What we have seen is that a family of its sectors \emph{does} possess a $U(1)$ gauge symmetry.  Mathematically, this is a non-invertible gauge symmetry implemented as projective isometries. Quantum simulations of models with such symmetries realize gauge theories through Hamiltonians that are not gauge-invariant themselves, provided that initial states are prepared in the sector with suitable gauge invariance.

It would be interesting to understand whether there are general mechanisms by which gauge invariance, with respect to some generic group $G$, can emerge in sectors of strongly fragmented systems. Furthermore, it remains an open question how one can characterize the different sectors of these models. Although we provided a novel mechanism to uniquely characterize many fragments of the $S=1$ dipole-conserving chain, this is not enough to characterize \emph{all} fragments. This suggests that there is a deeper structure to be uncovered, which could be of interest to better understand gauge theories and statistical physics alike. We leave these questions for future work.

\section{Acknowledgements}
We thank Debasish Banerjee, Andrea Bulgarelli, Klemen Kersic, Kiryl Pakrouski, Zlatko Papic, Yarden Scheffer, Arnab Sen, Tin Sulejmanpasic, and  Weronika Wiesiolek for enlightening discussions. This research was supported in part by grant NSF PHY-2309135 to the Kavli Institute for Theoretical Physics (KITP). MKM is grateful for the hospitality of Perimeter Institute where part of this work was carried out. Research at Perimeter Institute is supported in part by the Government of Canada through the Department of Innovation, Science and Economic Development and by the Province of Ontario through the Ministry of Colleges and Universities. This research was also supported in part by the Simons Foundation through the Simons Foundation Emmy Noether Fellows Program at Perimeter Institute.

\bibliographystyle{JHEP}
\bibliography{references}

\end{document}